\documentclass[12pt,psfig]{article}

\usepackage{graphicx}
\usepackage{amssymb}
\title{Hybrid atomistic-continuum methods for multiscale hydrodynamics}
\author{
Hettithanthrige S. Wijesinghe and Nicolas~G.~Hadjiconstantinou\\
{\it Mechanical Engineering Department}\\
{\it Massachusetts Institute of Technology}\\
{\it Cambridge, MA 02139} }

\date{\today}

\begin{document}

\maketitle

\begin{abstract}
We discuss hybrid atomistic-continuum methods for multiscale hydrodynamic 
applications. Both dense fluid and dilute gas formulations are considered.  
The choice of coupling method and its relation to the fluid physics is discussed. 
The differences in hybrid methods resulting from underlying compressible and 
incompressible continuum formulations as well as the importance of timescale 
decoupling are highlighted. We also discuss recently developed 
compressible and incompressible hybrid methods for dilute gases. The
incompressible framework is based on the Schwarz alternating method whereas
the compressible method is a multi-species, fully adaptive mesh and 
algorithm refinement approach which introduces the direct simulation Monte 
Carlo at the finest level of mesh refinement. 
\end{abstract}

\section{Introduction}
\label{introsection}
By limiting the molecular treatment to regions where it is needed, a hybrid 
method allows the simulation of complex thermo-fluid phenomena which require 
modeling at the microscale without the prohibitive cost of a fully molecular 
calculation. In what follows we provide an overview of this rapidly expanding 
field and discuss recent developments. We also present archetypal hybrid methods 
for incompressible and compressible flows; the hybrid method for incompressible gas 
flow is based on the Schwarz alternating coupling method which uses Chapman-Enskog 
boundary condition imposition, the hybrid method for compressible flow is the recently 
developed~\cite{imece2003} flux-coupling based, multispecies adaptive mesh and algorithm 
refinement scheme that extends adaptive mesh refinement by introducing the molecular 
description at the finest level of refinement.

Over the years a fair number of hybrid simulation frameworks has been proposed 
leading to some confusion over the relative merits and applicability of each approach. 
Original hybrid methods focused on dilute gases~\cite{ref:wadsworth90, ref:wadsworth92, 
ref:eggers, ref:hash95}, which are arguably easier to deal with within a hybrid framework 
than dense fluids, mainly because boundary condition imposition is significantly easier 
in gases. The first hybrid methods for dense fluids appeared a few years later 
\cite{thompson,me,jcp99,Flekkoy}. These initial attempts have led to a 
better understanding of the challenges associated with hybrid methods.

To a large extent, the two major issues in developing a hybrid method is the 
choice of a coupling method and the imposition of boundary conditions on the 
molecular simulation. Generally speaking, these two can be viewed as decoupled,
in the sense that the coupling technique can be developed on the basis of 
matching two compatible and equivalent over some region of space hydrodynamic 
descriptions and can thus be borrowed from the already existing and extensive 
continuum-based  numerical methods literature. The choice of coupling technique is further discussed in section \ref{coupling}. Boundary condition imposition 
can again be considered in a decoupled sense and can be posed as a general problem of imposing ``macroscopic'' boundary conditions on a molecular simulation. In our opinion, this is a very challenging problem that has not been, in general, resolved to date completely satisfactorily. Boundary condition imposition on the molecular sub-domain is discussed in section \ref{bcsection}. Boundary condition imposition on the continuum sub-domain is generally well understood, as is the process of extracting macroscopic fields from molecular simulations (typically achieved through averaging).

In section \ref{dsmcsec} we give a brief description of the direct simulation Monte Carlo (DSMC), the dilute gas simulation method used in this work. In section \ref{schwarz} we demonstrate a hybrid scheme suitable for low speed, incompressible gaseous flows based on the Schwarz alternating method.  This scheme introduces Chapman-Enskog boundary condition imposition in incompressible hybrid formulations. Subsequently, in section \ref{compressible} we discuss a recently developed \cite{imece2003} multi species compressible formulation for gases which introduces the molecular simulation as the finest level of refinement within a fully 
adaptive mesh refinement scheme. We finish with some concluding remarks.

\section{Developing a  hybrid method}
\subsection{The choice of coupling method}
\label{coupling}
Coupling a continuum to a molecular description is meaningful in a region where both can be presumed valid. In choosing a coupling method it is therefore convenient to draw upon the wealth of experience and large cadre of coupling methods nearly 50 years of continuum computational fluid dynamics have brought us. Coupling methods for the compressible and incompressible formulations generally differ, since the two correspond to two different physical and mathematical hydrodynamic limits. Faithful to their mathematical formulations, the compressible formulation 
lends itself naturally to time explicit flux-based (control-volume-type) coupling while incompressible formulations are typically coupled using either state properties (Dirichlet) 
or gradient information (Neumann). 

Given that the two formulations have different limits of applicability and physical regimes in which each is significantly
more  efficient than the other, care must be exercised when selecting the ingredients of the hybrid method. In other words, the choice of a coupling method and continuum sub-domain formulation  needs to be based on the degree on which compressibility effects are important in the problem of interest and not on a preset notion that a particular coupling method is more appropriate than all others. The latter approach was recently pursued in a variety of studies which enforce the use of a control-volume-type approach to steady and essentially incompressible problems to achieve coupling by time explicit flux matching. This approach is not recommended. On the contrary, for an efficient simulation method, similarly to the case of continuum solution methods, it is important to allow the flow {\it physics} to dictate the appropriate formulation, while the numerical implementation is chosen to cater to the particular requirements of the latter. Below, we expand on some of the considerations which influence the choice of coupling method under the assumption that the hybrid method is applied to problems of {\it practical interest} and therefore the continuum subdomain is appropriately large. Our discussion focuses on timescale considerations that are more complex but equally important to limitations resulting from lengthscale considerations, such as the size of the molecular region(s). 

It is well known \cite{wesseling} that the timestep for explicit integration of the compressible formulation, $\tau_c$, scales with the physical timestep of the problem, $\tau_{\Delta x} $($=\Delta x/U$, where $\Delta x$ is the numerical grid spacing and $U$ is the characteristic velocity), 
according to
\begin{equation}
\tau_c\leq \frac{M}{1+M} \tau_{\Delta x}
\label{mach}
\end{equation}
where $M$ is the Mach number. As the Mach number becomes small, we are faced with the well-known stiffness problem whereby the numerical efficiency of the solution method suffers \cite{wesseling} because the compressible formulation resolves the acoustic modes when those are not important. For this reason, when the Mach number is small, the incompressible formulation is used which allows integration at the physical timestep $\tau_{\Delta x}$. 
In the hybrid case matters are complicated by the introduction of  the molecular integration
timestep, $\tau_m$, which is at most of the order of $\tau_c$ (in some cases in gases when $\Delta x \leq \lambda$, where $\lambda$ is the molecular mean free path) and in most cases significantly smaller. 
One consequence of equation (\ref{mach}) is that as the global domain of interest grows, the total integration time grows, and transient calculations in which the molecular subdomain is explicitly integrated in time become more computationally expensive and eventually infeasible. The severity of this problem increases with decreasing Mach number and makes unsteady incompressible problems very computationally expensive. New integrative frameworks which coarse grain the time integration of the molecular subdomain are therefore required. 

Fortunately, for low speed steady problems implicit (iterative) methods exist which provide solutions without the need for explicit integration of the molecular domain to the global problem steady state.   The particular method used here is known as the Schwarz method and is discussed further in Section \ref{schwarz}. This method decouples the global evolution timescale from the molecular
evolution timescale (and timestep) by achieving  convergence to the global problem steady state through an iteration between steady state
solutions of the continuum and molecular subdomains. Because the molecular subdomain is small, explicit integration to its steady state is feasible. Although the steady assumption may appear restrictive, it is interesting to note that the vast majority of  both compressible and incompressible test problems solved to date and all incompressible practical problems of interest \cite{aluru} solved by hybrid methods have been steady. A variety of other iterative methods may be
suitable as they provide for timescale decoupling. The choice of the Schwarz coupling method, which uses state variables instead of fluxes to achieve matching,  was motivated by the fact (as explained below) that state variables suffer from smaller statistical noise and are thus easier to prescribe on a continuum formulation.

The above observations do not preclude the use of the compressible formulation in the continuum subdomain for low speed flows. In fact, preconditioning techniques which allow the use of the compressible formulation at very low Mach numbers have been developed \cite{wesseling}. Such a formulation can, {\it in principle}, be used to solve the continuum sub-problem while this is being coupled to the molecular sub-problem via an implicit (eg. Schwarz) iteration. What should be avoided is a time-explicit control-volume-type coupling procedure for solving essentially incompressible steady state problems.

The issues discussed above have not been very apparent to date because in typical test problems published so far, the continuum and atomistic subdomains are of the same size (and, of course, small). In this case the large cost of the molecular subdomain masks the cost of the continuum subdomain  and also  typical evolution timescales (or times to steady state) are small. It should not be forgotten, however, that hybrid methods make sense when the continuum subdomain is significantly larger than the molecular subdomain.

Although the continuum subdomain stiffness in the limit $M\rightarrow 0$ (see eq (\ref{mach})) may be remedied by implicit timestepping methods \cite{yuan} or preconditioning approaches, control-volume-based hybrid approaches suffer from adverse signal to noise ratios in connection with the averaging required for imposition of boundary conditions from the molecular sub-domain to the  continuum sub-domain. In the case of an ideal 
gas (where compressible formulations are typical) it has been shown in \cite{flucterror} that, for the same number of samples, flux (shear stress, heat flux) 
averaging exhibits relative noise, $E_f$, which scales as 
\begin{equation}
E_f\sim\frac{E_{sv}}{Kn}
\end{equation}
where $E_{sv}$ is the relative noise in the corresponding state variable 
(velocity, temperature). Here $Kn=\lambda/L$ is the Knudsen number based on the characteristic lengthscale of the transport gradients, $L$, and  $\lambda$ is the mean free path which is expected to be much smaller than $L$ since, by assumption, a continuum sub-domain is present. It thus appears that flux coupling will be significantly disadvantaged in this case, since the number of samples required to make $E_f\approx E_{sv}$ scales as $1/Kn^2$ times the number of samples required by state-variable averaging. On the other hand, Schwarz-type iterative methods based on the incompressible physics of the flow require a fair number of iterations for convergence ($O(10)$). These iterations require the re-evaluation of the molecular solution. This is an additional computational cost that is not shared by control-volume-type  or explicit incompressible approaches. At this time, the choice between a time-explicit formulation or an iterative (Schwarz-type) iteration for incompressible unsteady problems is not clear  and may be problem dependent. Despite the fact that as L grows the advantage seems to shift towards iterative methods, we should recall that from equation (\ref{mach}), unless time coarse-graining techniques are developed, large, low-speed,  unsteady problems are currently too expensive to be feasible by either approach.

\subsection{Boundary condition imposition}
\label{bcsection}
\begin{figure}
 \begin{center}
 \includegraphics[height=2.5 in]{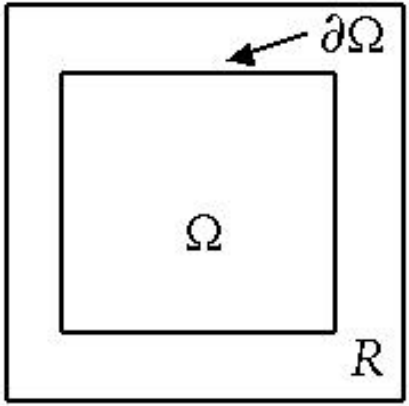}
 \caption{Continuum to atomistic boundary condition imposition using reservoirs.}
 \label{reservoirfig}
 \end{center}
\end{figure}

Consider the molecular region $\Omega$ on the boundary of which, 
$\partial \Omega$, we wish to impose a set of hydrodynamic (macroscopic) 
boundary conditions. Typical implementations require the use of particle 
reservoirs ${\cal R}$ (see Fig \ref{reservoirfig}) in which particle dynamics 
may be altered in such a way that the desired boundary conditions appear on   
$\partial \Omega$; the hope is that the influence of the perturbed dynamics 
in the reservoir regions decays sufficiently fast and does not propagate into the region of interest, that is, the relaxation distance both for the velocity 
distribution function and the fluid structure is small compared to the characteristic size of $\Omega$. 

In a dilute gas, the non-equilibrium distribution function in the continuum limit 
has been characterized \cite{chapman} and is known as the Chapman-Enskog distribution. 
Use of this distribution to impose boundary conditions on molecular simulations of 
dilute gases results in a robust, accurate and theoretically elegant approach. 
Typical implementations \cite{alex} require the use of particle 
generation and initialization within ${\cal R}$. Particles 
that move into $\Omega$ within the simulation timestep are added to the simulation whereas particles remaining in ${\cal R}$ are discarded. See chapter \ref{schwarz} for details.

Unfortunately, for dense fluids where not only the particle velocities but also 
the fluid structure is important and needs to be imposed, no theoretical results 
for their distributions exist.  A related issue is that of domain termination; 
due to particle interactions, $\Omega$, or in the presence of a reservoir,
${\cal R}$, needs to be terminated in a way 
that does not have a big effect on the fluid state inside of $\Omega$.

As a result, researchers 
have experimented with possible methods to impose boundary conditions. It is 
now known that similarly to a dilute gas, use of a Maxwell-Boltzmann 
distribution for the velocities leads to slip \cite{me}. 
Li et al. \cite{li} used a Chapman-Enskog distribution to impose boundary conditions 
to generate a dense-fluid shear flow. In this approach, particles crossing 
$\partial \Omega$ acquire velocities that are drawn from a Chapman-Enskog 
distribution parametrized by the local values of the required velocity and stress 
boundary condition. Although this approach was only tested for a Couette flow, it 
appears to give reasonable results (within molecular fluctuations). Because in 
Couette flow no flow normal to $\partial \Omega$ exists,  $\partial \Omega$ 
can be used as symmetry boundary separating two back-to-back shear flows; this 
sidesteps the issue of domain termination.

In a different approach, Flekkoy et al. \cite{Flekkoy} use external forces to 
impose boundary conditions. More specifically, in the reservoir region they apply 
an external field of such magnitude that the total force on the fluid particles 
in the reservoir region is the one required by momentum conservation. 
They then terminate their reservoir region by using an ad-hoc weighing 
factor for the distribution of this force on particles within ${\cal R}$ that prevents 
particles from leaving the reservoir region. In particular, they chose a weighing 
factor that diverges as particles approach the edge of ${\cal R}$ such that 
particles do not escape the reservoir region while particles introduced there move 
towards $\Omega$. Particles introduced into the reservoir are given 
velocities drawn from a Maxwell-Boltzmann distribution, while a Langevin thermostat 
keeps the temperature constant. The method appears to be successful although 
the non-unique (ad-hoc) choice of force fields and Maxwell-Boltzmann distribution makes 
it not very theoretically pleasing. It is also not clear what the effect of 
these forces are on the local fluid state (it is well known that even in a dilute gas 
\cite{gravity} gravity driven flow exhibits significant deviations from Navier-Stokes behavior) but this 
effect is probably negligible since force fields are only acting in the reservoir region. 
Delgado-Buscalioni and Coveney~\cite{ref:delgado} refined the above approach by using
an Usher algorithm to insert particles in the energy landscape such that they have the
desired specific energy, which is beneficial to imposing a desired energy current
while eliminating the risk of particle overlap at some computational cost. This approach
uses a Maxwell--Boltzmann distribution however for the initial velocities of the 
inserted particles. Temperature gradients are imposed by a small number of thermostats
placed in the direction of the gradient. Although no proof exists that the disturbance
to the particle dynamics is small, it appears that this technique is successful at imposing
boundary conditions with moderate error. Boundary conditions on MD simulations
can also be imposed through the method of constraint dynamics \cite{thompson}. Although the approach in \cite{thompson} did not allow hydrodynamic fluxes across the matching interface, this feature can be integrated into this approach with a suitable domain termination.

A method for terminating molecular dynamics simulations with small effect on 
particle dynamics has been suggested and used in \cite{me}. This simply 
involves making the reservoir region fully periodic. In this manner, the boundary conditions on  $\partial \Omega$ also impose a boundary value problem 
on ${\cal R}$, where the inflow to $\Omega$ is the outflow from ${\cal R}$. 
As  ${\cal R}$ becomes bigger, the gradients in ${\cal R}$ become smaller and thus the flowfield in ${\cal R}$ will have a small effect on the solution in  $\Omega$. The disadvantage of 
this method is the number of particles that are needed to fill  ${\cal R}$  as 
this grows, especially in high dimensions.

We believe that significant contributions can still be made by developing methods to impose boundary conditions in hydrodynamically consistent and, most importantly, rigorous approaches. 

\section{The direct simulation Monte Carlo}
\label{dsmcsec}
The DSMC method was proposed by Bird \cite{bird} in the 1960's and has been used 
extensively to model rarefied gas flows. A comprehensive discussion of DSMC can be 
found in the review article by Alexander et. al.~\cite{dsmcintro}. The DSMC algorithm is based on 
the assumption that a small number of representative ``computational particles'' can accurately 
capture the hydrodynamics of a dilute gas as given by the Boltzmann equation. 
Air under standard conditions narrowly meets the dilute gas criterion.
Empirical results \cite{bird} show that a small number ($\approx 20$) 
of computational particles per 
cubic molecular mean free path is sufficient
to capture the relevant physics. This is approximately 2 orders of magnitude
smaller than the actual number of gas atoms/molecules contained in the same volume. 
This is one source of DSMC's significant computational advantage over a fully molecular
simulation. 

DSMC solves the Boltzmann equation using a splitting approach: the time 
evolution of the system is approximated by a sequence of discrete timesteps, $\Delta t$, in which particles undergo successively collisionless advection
and collisions. Collisions are performed between randomly chosen particle pairs
within small cells of linear  size $\Delta x$. The flow solution is determined by 
averaging the individual particle properties over space and time. This approach 
has been shown to produce correct solutions of 
the Boltzmann equation in the limit $\Delta x$, $\Delta t\rightarrow 0$ \cite{Wagner:Proof}. 
The splitting approach eliminates the computational cost
associated with integrating the equations of motion 
of all particles, but most importantly allows the timestep to be
significantly larger (see also below) than a typical timestep in a hard sphere 
molecular dynamics simulation. This is another reason why DSMC is significantly 
more computationally efficient than ``brute force'' molecular dynamics. 

Recent studies \cite{timeerror1,timeerror2} have shown that for steady flows, or 
flows which are evolving at timescales that are long compared to the molecular 
relaxation times, a finite timestep leads to a truncation error that manifests 
itself in the form of timestep-dependent transport coefficients; this error
has been shown to be of the order of 5\% when the timestep is of the order 
of a mean free time and goes to zero as $\Delta t^2$. Quadratic dependence of
transport coefficients on the collision cell size $\Delta x$ was shown in 
\cite{cellerror}.

\section{The Schwarz method for incompressible formulations}
\label{schwarz}
Although in some cases compressibility may be important, a large number of 
applications are typically characterized by flows where use of the
incompressible formulation results in a significantly more efficient approach
\cite{wesseling}. As explained in the introduction section, our definition of
incompressible formulation is based on the {\it flow physics} and not on the
numerical method used. Although we have used here a finite element discretization based on the incompressible formulation, we believe that a preconditioned
compressible formulation could also be used to solve the continuum subdomain problem if it could be successfully matched to the molecular solution through a coupling method which takes into account the elliptic nature of the (low speed) problem to provide solution matching consistent with the flow physics.

Here, matching is achieved through an iterative procedure based on the Schwarz alternating method for the treatment of steady-state problems.
The Schwarz method was 
originally proposed for molecular dynamics-continuum methods in \cite{me} and extended in \cite{jcp99}, but it is equally 
applicable to DSMC-continuum hybrid methods \cite{aluru,sanithrgd23}. 
This approach was chosen because of its ability to couple different descriptions 
through Dirichlet boundary conditions (easier to impose on dense-system 
molecular simulations compared to flux conditions, because fluxes are non-local 
in dense systems), and its ability to reach the solution steady state in an 
implicit manner. The importance of the latter characteristic cannot be overemphasized; 
the implicit convergence in time guarantees timescale decoupling that is necessary 
for the solution of macroscopic problems; the integration of molecular trajectories 
at the molecular timestep for total times corresponding to macroscopic evolution 
times is, and will for a long time be, infeasible. 

Within the Schwarz coupling framework, an overlap region facilitates information exchange 
between the continuum and atomistic subdomains in the form of Dirichlet boundary 
conditions. A steady state continuum solution is first obtained using boundary 
conditions taken from the atomistic subdomain solution. For the first iteration 
this latter solution can be a guess. A steady state atomistic solution is 
then found using boundary conditions taken from the continuum subdomain. 
This exchange of boundary conditions corresponds to a single Schwarz iteration. 
Successive Schwarz iterations are repeated until convergence, i.e. until the solutions 
in the two subdomains are identical in the overlap region. 

The Schwarz method was recently applied \cite{aluru} to the simulation of flow 
through micromachined filters. These filters have passages that are sufficiently 
small that require a molecular description for the simulation of the flow through 
them. Depending on the geometry and number of filter stages the authors have 
reported computational savings ranging from 2 to 100. The approach in \cite{aluru} 
used a Maxwellian velocity distribution and a ``control mechanism'' to 
impose the flowfield on the molecular simulation. This approach, although succesful 
in quasi one-dimensional flows, is not very general; additionally, it is well known 
that using a Maxwellian distribution to impose hydrodynamic boundary conditions, 
in general, if uncorrected will lead to slip (discrepancy between the
imposed and observed boundary conditions). 
General boundary condition imposition on dilute-gas 
molecular simulations can be performed  using the Chapman-Enskog velocity 
distribution \cite{alex, chapman}. This approach eliminates the need for a
feedback correction since supplying the correct local distribution function eliminates slip. A Chapman-Enskog procedure for the Schwarz method is described 
below. Extensions of the Schwarz method to time-dependent problems is currently
under investigation \cite{sanith}, although, as shown by equation (1), when the Mach number is low, the disparity between the molecular and hydrodynamic timescales makes this a very stiff problem.

\subsection{Driven cavity test problem}

\begin{figure}
 \begin{center}
 \includegraphics[height=3.5 in]{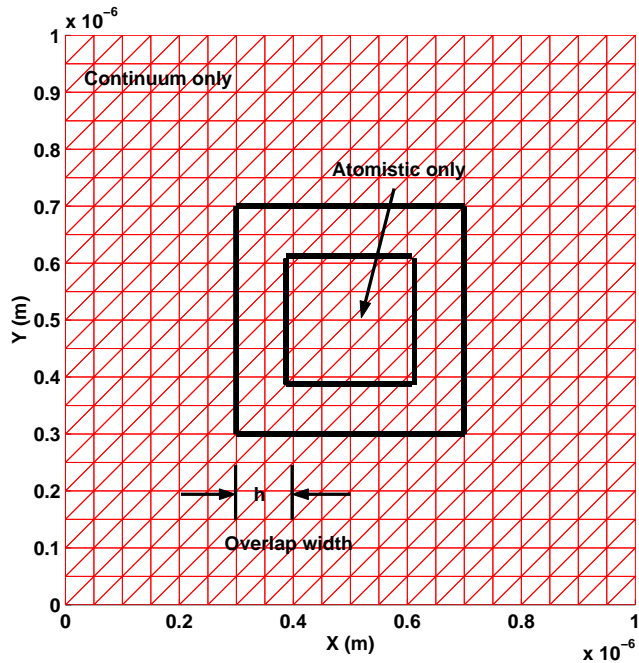}
 \caption{Continuum and atomistic sub-domains for Schwarz coupling in two--dimensions.}
 \label{fig:domain}
 \end{center}
\end{figure}

In this section we discuss the Schwarz alternating method in the 
context of the solution of the driven cavity problem. We pay particular attention
to the imposition of boundary conditions on the DSMC domain using a Chapman-Enskog distribution which is arguably the most rigorous and general approach. For 
illustration and verification purposes we solve the steady driven cavity problem (see Figure~\ref{fig:domain}), in 
which the continuum subdomain is described by the Navier-Stokes equations solved
by finite element discretization. The hybrid solution is expected to recover
the fully continuum solution since the atomistic subdomain is far from solid
boundaries and from regions of large velocity gradients. This test therefore 
provides a consistency check for the scheme.

Standard Dirichlet velocity boundary 
conditions for a driven cavity problem were applied on the continuum subdomain; 
the $u$ velocity component on the left, right and lower walls were held at zero 
while the upper wall $u$ velocity was set to $50$ m/s. The $v$ velocity component 
on all boundaries was set to zero. Despite the high velocity, the flow is essentially
incompressible and isothermal. The pressure is scaled by setting the middle 
node on the lower boundary at atmospheric pressure ($1.013\times 10^5$ Pa).

\begin{figure}
 \begin{center}
 \includegraphics[height=3.5 in]{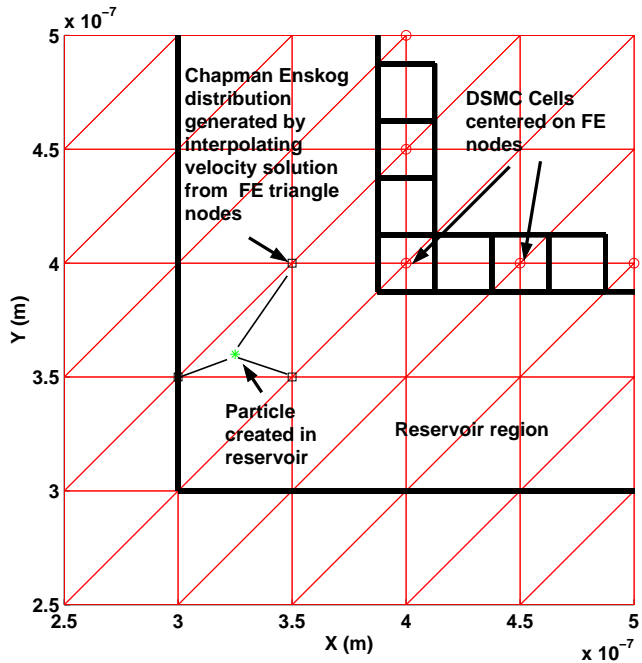}
 \caption{Particle reservoir in the overlap region.}
 \label{fig:interpolations}
 \end{center}
\end{figure}

The imposition of boundary conditions on the atomistic 
subdomain is facilitated by a particle reservoir as 
shown in Figure~\ref{fig:interpolations}. Particles are created at locations 
$x,y$ within the reservoir with velocities $C_x,C_y$ drawn from a Chapman-Enskog 
velocity distribution. The Chapman Enskog distribution is
generated \cite{ref:garcia98} 
by using the mean and gradient of velocities from the 
 continuum solution, that is, the number and spatial distribution of particles in the reservoir are chosen 
according to the overlying continuum cell mean density and density gradients. 
After particles are created in the reservoir they move for a single 
DSMC timestep. Particles that enter DSMC cells are incorporated into the 
standard convection/collision routines of the DSMC algorithm. Particles that 
remain in the reservoir are discarded. Particles that leave the DSMC 
domain are also deleted from the computation. 

\begin{figure}
 \begin{center}
 \includegraphics[height=3.5 in]{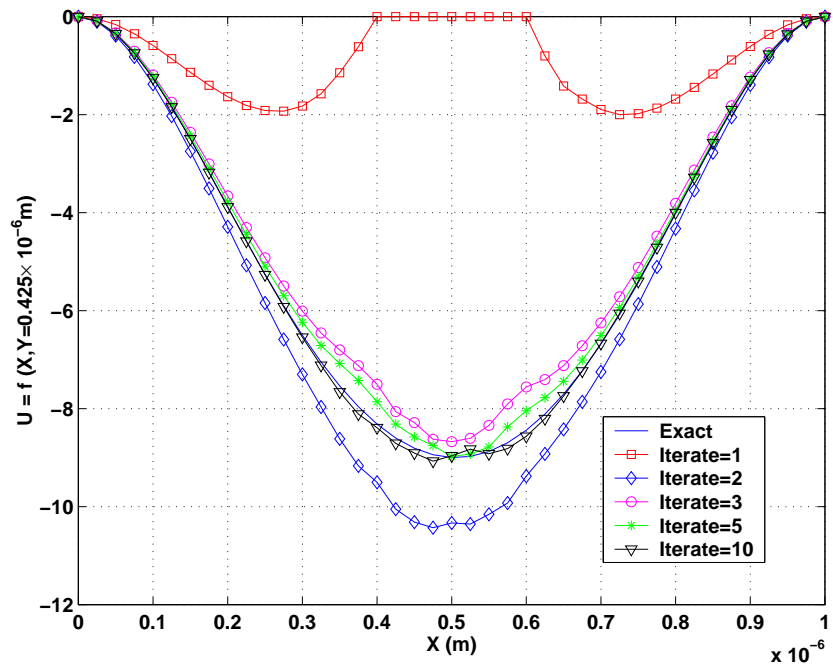}
 \caption{Convergence of the $u$ velocity component along 
 the $y=0.425\times 10^{-6}m$ plane with successive Schwarz iterations.}
 \label{fig:u_convergence}
 \end{center}
\end{figure}

The rapid convergence of the Schwarz approach is demonstrated in 
Figure~\ref{fig:u_convergence}. The continuum numerical solution is reached to within 
$\pm 10\%$ at the 3rd Schwarz iteration and to within $\pm 2\%$ at the 10th 
Schwarz iteration. Our error estimate which includes the effects of statistical noise \cite{flucterror} and
discretization error due to finite timestep and cell size is approximately 2.5\%. Similar convergence of the $v$ velocity field is also observed. 

The close agreement with the fully continuum results indicates that the Chapman-Enskog
procedure is not only theoretically appropriate but also robust. Despite a Reynolds
number of $Re\approx1$, the Schwarz method (originally only shown to converge
for elliptic problems \cite{lions}) converges with negligible error. This is in agreement with
the findings of Liu \cite{liu} who has recently shown that the Schwarz method is expected to
converge for $Re \sim O(1)$.

\section{Adaptive mesh and algorithm refinement for compressible formulations}
\label{compressible}
As discussed above, consideration of the compressible equations of motion 
leads to hybrid methods which differ significantly from their incompressible 
counterparts. The hyperbolic nature of compressible flows  means that
steady state formulations typically do not offer a significant computational advantage, and as a result, explicit time integration is the preferred solution method and flux matching is the preferred coupling method. Given that the characteristic evolution time, $\tau_h$, scales with the system size, the largest problem  that can be captured by a hybrid method is limited by the separation of scales between the molecular integration time and $\tau_h$. Local mesh refinement techniques \cite{alex,imece2003} minimize the regions of space that need to be integrated at small CFL timesteps (due to a fine mesh), such as the regions adjoining the molecular subdomain. Implicit timestepping methods \cite{yuan} can also be used to speed up the time integration of the continuum subdomain. Unfortunately, although both approaches enhance the computational efficiency of the continuum sub-problem, they do not alleviate
the issues arising from the disparity between the molecular timestep and the
total integration time. 

As explained in the introduction, overwhelming computational costs can be incurred when using a control-volume-type approach to capture steady phenomena where compressibility effects are negligible as is in most cases in dense fluids. In this case the 
integration timestep of the 
continuum subdomain also becomes of the order of the molecular timescale, while
the continuum subdomain is, presumably, much larger than the molecular subdomain and evolves at a much longer timescale. For an example, see in \cite{Flekkoy}, where a 
compressible formulation for liquid flow results in a CFL timestep of 0.17 $\tau_{LJ}$ where $\tau_{LJ}\approx O(10^{-12}\mathrm{s})$ is the Lennard-Jones timescale. This appears to not have been fully appreciated 
by various groups which have attempted to develop dense-fluid hybrid methods based on the compressible continuum formulation and control-volume-type matching procedures to solve steady and essentially incompressible problems.

In hybrid continuum-DSMC methods, locally refining the continuum solution cells 
to the size of DSMC cells leads to a particularly seamless hybrid formulation 
in which DSMC cells differ from the neighboring continuum cells only by the fact 
that they are inherently fluctuating. (The DSMC timestep required for accurate 
solutions--see \cite{cellerror,timeerror1,timeerror2}--is very similar to the 
CFL timestep of a compressible formulation.) Coupled to a mesh refinement method,
this approach can tackle true multiscale phenomena as shown below.

Another characteristic inherent to compressible formulations is the possibility of 
describing parts of the domain by the Euler equations of motion \cite{imece2003}. 
In that case, consistent coupling to the molecular formulation can be performed 
using a Maxwell-Boltzmann distribution \cite{alex}.

In a recent paper \cite{tarta}, Alexander et al. have shown that explicit 
time-dependent flux-based formulations preserve the fluctuating nature of the 
molecular description within the molecular regions but the fluctuation amplitude 
decays rapidly within the continuum regions; correct fluctuation spectra can be 
obtained in the entire domain by solving a fluctuating hydrodynamics formulation 
\cite{landau} in the continuum sub-domain.

\subsection{Fully adaptive mesh and algorithm refinement for a dilute gas}
\label{amar}
The compressible formulation of Garcia et al.~\cite{alex}, referred to as AMAR 
(Adaptive Mesh and Algorithm Refinement), pioneered the use of mesh refinement 
as a natural framework for the introduction of the molecular description in a 
hybrid formulation. In AMAR the typical continuum mesh refinement capabilities 
are supplemented by an algorithmic refinement (continuum to atomistic) based on 
continuum breakdown criteria. This seamless transition is both theoretically and 
practically very appealing. 

In what follows we briefly discuss a recently developed \cite{imece2003,asme2003} 
{\it fully adaptive} AMAR method. In this method DSMC provides an atomistic description 
of the flow while the compressible two--fluid Euler equations serve as the 
continuum--scale model. The continuum and atomistic representations are coupled 
by matching fluxes at the continuum--atomistic interfaces and by proper averaging 
and interpolation of data between scales. This is performed 
in three steps; a) the continuum solution values are interpolated 
to create DSMC particles in the reservoir region, here called buffer cells, 
b) the conserved quantities 
in each continuum cell overlaying the DSMC region are replaced by averages 
over particles in the same region and c) fluxes recorded when particles 
cross the DSMC interface are used to correct the continuum solution in 
cells adjacent to the DSMC region. This coupling procedure 
makes the DSMC region appear as any other level in an AMR grid hierarchy. 
Similarly to the overlap region described for the Schwarz method
above, the Euler solution information is passed to the particles via buffer 
cells surrounding the DSMC region. At the beginning of each DSMC integration 
step, particles are created in the buffer cells using the continuum 
hydrodynamic values. 

The above algorithm allows grid and algorithm refinement based on any combination
of flow variables and their gradients. Density gradient based refinement has 
has been found to be generally robust and reliable. Concentration gradients or 
concentration values within some interval are also effective refinement criteria especially for multi-species flows involving concentration interfaces. 
In this particular implementation, refinement is triggered by spatial gradients  exceeding user defined tolerances. This approach follows from the continuum 
breakdown parameter method proposed by Bird~\cite{Bird:Breakdown}. 

Using the AMR capabilities provided by the Structured Adaptive Mesh Refinement 
Application Infrastructure (SAMRAI) developed at the Lawrence Livermore National 
Laboratory \cite{samraiblah}, the above adaptive framework has been implemented 
in a fully three-dimensional, massively parallel form in which, multiple 
molecular (DSMC) patches can be introduced or removed as needed. 

\begin{figure}
 \begin{center}
 \includegraphics[height=3.5in, angle=-90]{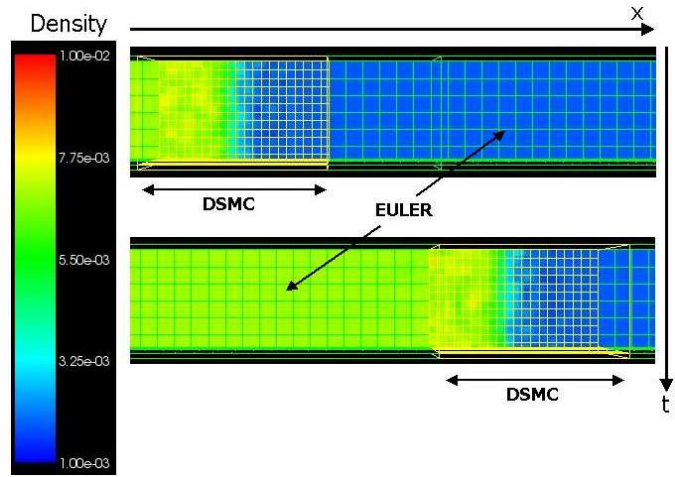}
 \caption{Moving Mach 10 shock wave though Argon.  The AMAR algorithm tracks the
 shock by adaptively moving the DSMC region with the shock front. }
 \label{fig:moving-shock}
 \end{center}
\end{figure}

\begin{figure}
 \begin{center}
 \includegraphics[height=3.5in]{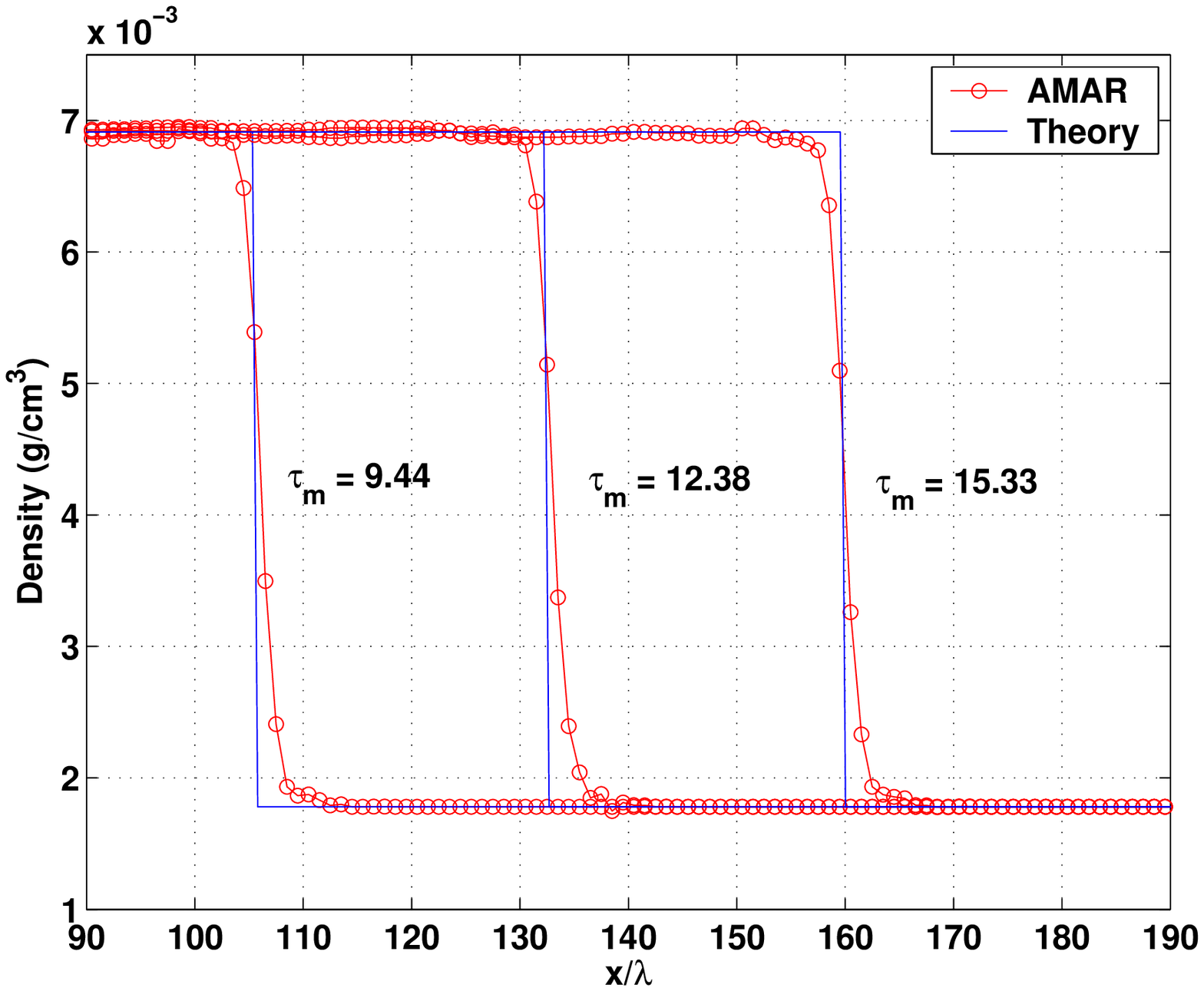}
 \caption{Moving Mach 10 shock wave though Argon. The AMAR profile 
 (red dots) is compared with the analytical time evolution of the initial 
 discontinuity (blue lines). $\tau_m$ is the mean collision time.  }
 \label{fig:shock-profile}
 \end{center}
\end{figure}

Figure~\ref{fig:moving-shock} shows the adaptive tracking of a shockwave of
Mach number 10 used as a validation test for this method. Density gradient based mesh 
refinement ensures the DSMC region tracks the shock front accurately.
Furthermore, as shown in Figure~\ref{fig:shock-profile} the density profile
of the shock wave remains smooth and is devoid of oscillations that are known
to plague traditional shock capturing schemes~\cite{Aroroa:Roe, Woodward:Colella}.

\section{Discussion}
\label{discussionsection}
One of the most important messages of this paper is that boundary condition imposition on molecular domains is quite independent of the choice of solution coupling approach. As a example, consider the Schwarz method which provides a recipe for making solutions in various subdomains globally consistent subject to exchange of Dirichlet conditions. The imposition of these boundary conditions can be achieved through any method and no certain method is favored by the coupling approach. Flexibility in adopting {\it appropriate  elements} from previous approaches, and the importance of chosing the coupling method according to the flow physics are  key steps to the development of more sophisticated,  next-generation hybrid methods. 

Although hybrid methods provide significant savings by limiting molecular 
solutions only to the regions where they are needed, solution of time-evolving 
problems which span a large range of timescales is still not possible if the 
molecular domain, however small, needs to be integrated for the total time of interest. New frameworks are therefore required which allow timescale decoupling or coarse  grained time evolution of molecular simulations. 

Significant computational savings can be obtained by using the incompressible 
formulation when appropriate for steady problems. 
Neglect of these simplifications
can lead to a problem that is simply intractable when the continuum subdomain is appropriately large. It is interesting to note that, when a hybrid method was used to solve a problem of practical interest \cite{aluru} while providing computational savings, the Schwarz method was preferred because it provides a steady solution framework with timescale decoupling.

For dilute gases the Chapman-Enskog distribution provides a robust and accurate
method for imposing boundary conditions. Further work is required for the development
of similar frameworks for dense liquids.

\section{Acknowledgements}
The authors wish to thank R. Hornung and A. L. Garcia for help with the 
computations and valuable comments and discussions, and A. T. Patera and B. J. Alder 
for helpful comments and discussions. This work was supported in part by the 
Center for Computational Engineering, and the Center for Advanced Scientific 
Computing, Lawrence Livermore National Laboratory, US Department of Energy, 
W-7405-ENG-48. The authors also acknowledge the financial support from the 
University of Singapore through the Singapore-MIT alliance.

\end{document}